\documentclass[runningheads]{llncs}
\usepackage{graphicx}
\usepackage{color}
\begin{document}
\title{Beyond attention: deriving biologically interpretable insights from weakly-supervised multiple-instance learning models}
\titlerunning{Beyond attention}
%
\author{
Willem Bonnaff\'e\inst{1,2,*} \and
CRUK ICGC Prostate Group \inst{3,a} \and
Freddie Hamdy\inst{2,a} \and
Yang Hu\inst{1,4,a} \and
Ian Mills\inst{2,a} \and
Jens Rittscher\inst{1,3,5,a} \and
Clare Verrill\inst{2,a} \and
Dan J. Woodcock\inst{2,a}
}
\authorrunning{W. Bonnaff\'e et al.}
%
\institute{
Big Data Institute, Li Ka Shing Center for Health Information and Discovery, University of Oxford, Oxford, UK. \and
Nuffield Department of Surgical Sciences, University of Oxford, Oxford, UK. \and
A list of members and affiliations appears in the acknowledgements. \and
Nuffield Department of Medicine, University of Oxford, Oxford, UK. \and
Department of Engineering Science, University of Oxford, Oxford, UK. \\
$^*$ Corresponding author (willem.bonnaffe@nds.ox.ac.uk). \\
$^a$ Authors listed in alphabetical order.
}
\maketitle              
\begin{abstract}

Recent advances in attention-based multiple instance learning (MIL) have improved our insights into the tissue regions that models rely on to make predictions in digital pathology. However, the interpretability of these approaches is still limited. In particular, they do not report whether high-attention regions are positively or negatively associated with the class labels or how well these regions correspond to previously established clinical and biological knowledge. We address this by introducing a post-training methodology to analyse MIL models. Firstly, we introduce prediction-attention-weighted (PAW) maps by combining tile-level attention and prediction scores produced by a refined encoder, allowing us to quantify the predictive contribution of high-attention regions. Secondly, we introduce a biological feature instantiation technique by integrating PAW maps with nuclei segmentation masks. This further improves interpretability by providing biologically meaningful features related to the cellular organisation of the tissue and facilitates comparisons with known clinical features. We illustrate the utility of our approach by comparing PAW maps obtained for prostate cancer diagnosis (i.e. samples containing malignant tissue, 381/516 tissue samples) and prognosis (i.e. samples from patients with biochemical recurrence following surgery, 98/663 tissue samples) in a cohort of patients from the international cancer genome consortium (ICGC UK Prostate Group). Our approach reveals that regions that are predictive of adverse prognosis do not tend to co-locate with the tumour regions, indicating that non-cancer cells should also be studied when evaluating prognosis.

\end{abstract}
\section{Introduction}

Pathologists rely on specific morphological patterns to diagnose cancer, such as nuclei density, or glandular shape in the case of prostate cancer (PCa), which are derived from our mechanistic understanding of tumour histopathology \cite{Humphrey2004}. 
Prognosis in PCa, i.e. prediction of the likely outcome of a disease, is measured by biochemical recurrence (BCR), which corresponds to the rise in prostate-specific antigen (PSA) levels after treatment \cite{Pinckaers2022}. 
The association between BCR and histological features is not clear \cite{Pinckaers2022}, though there is increasing evidence of association with tumour-adjacent regions of the stroma \cite{Finak2008,Wu2016,Lee2017,Kemi2018,Gonzalez2022}.

The identification of interpretable morphological features predictive of cancer progression has become a main focus in computational pathology, as this can help better stratify patient into high- or low-risk cases in first-line diagnostic and avoid under- or over-treatment \cite{Pinckaers2022}.
Current deep learning approaches are often seen as unsuitable for the discovery of such interpretable features \cite{Rudin2018}, for lack of a methodology that translates abstract features produced by trained models into biologically meaningful properties of the tissue.
Recent multi-instance learning (MIL) approaches \cite{Ghaffari2022,Schrammen2022}, such as CLAM \cite{Lu2021}, include attention-based learning to find most informative regions of the images for the predictions.
Yet, attention is not enough, as it does not reveal whether these regions have a negative or positive association with the predictions of the model \cite{Rudin2018,Serrano2019}.
For instance, in the context of cancer diagnosis, which involves classifying benign vs tumour tissue, the attention maps would indiscriminately identify areas that associate with either benign or tumour tissue, given that both are useful evidence to establish whether the sample is overall benign or tumour tissue.

In this work, we propose a post-training interpretation methodology for MIL models, through \textit{prediction-attention coupling} and \textit{biological feature instantiation}. 
Prediction-attention coupling corresponds to coupling attention with tile-level prediction scores into prediction-attention-weighted (PAW) maps to identify regions that are predictive of the outcome. 
For instance, in the context of tumour diagnosis, PAW maps would indicate whether high-attention areas are more benign- or tumour-predictive, depending on the tile-level prediction scores.
Biological feature instantiation can then be used to derive biologically meaningful features from highly predictive regions. 
For the sake of simplicity, we focus here on local nuclei density, as it is a property of the tissue commonly used by pathologist, and because nuclei features have been be linked to BCR \cite{Lee2017}.

We introduce our methodology as a post-training step in the self-interactive MIL framework (Inter-MIL)~\cite{Hu2022}, which was developed to enhance the quality of the embeddings by iteratively training the tile-level encoder by using high-attention tiles identified by a slide-level classifier.
Inter-MIL is a suitable basis for our methodology as it provides higher quality features than standard MIL \cite{Hu2022}, tailored for the specific outcome that we are trying to predict, and generates prediction scores at the tile-level.
The advantage of implementing our methodology as a post-training step is that it does not require re-training the models nor limit the range of abstract features that the model can learn.
This makes it possible to efficiently explore a range of biologically informative features and simultaneously avoid a bias in the training of the model that may be caused by focusing on an arbitrarily chosen set of interpretable features. The potential impact of the proposed methodology is discussed in the context of samples from an extensive prostate cancer cohort. 
Specifically, we demonstrate that regions that are predictive of diagnosis in PCa differ from regions predictive of BCR in their localisation in the tissue and in their local cell density, and that attention alone is not enough to reveal these differences. 

\section{Methods}

\begin{figure}[t]
\centering
\includegraphics[width=\textwidth]{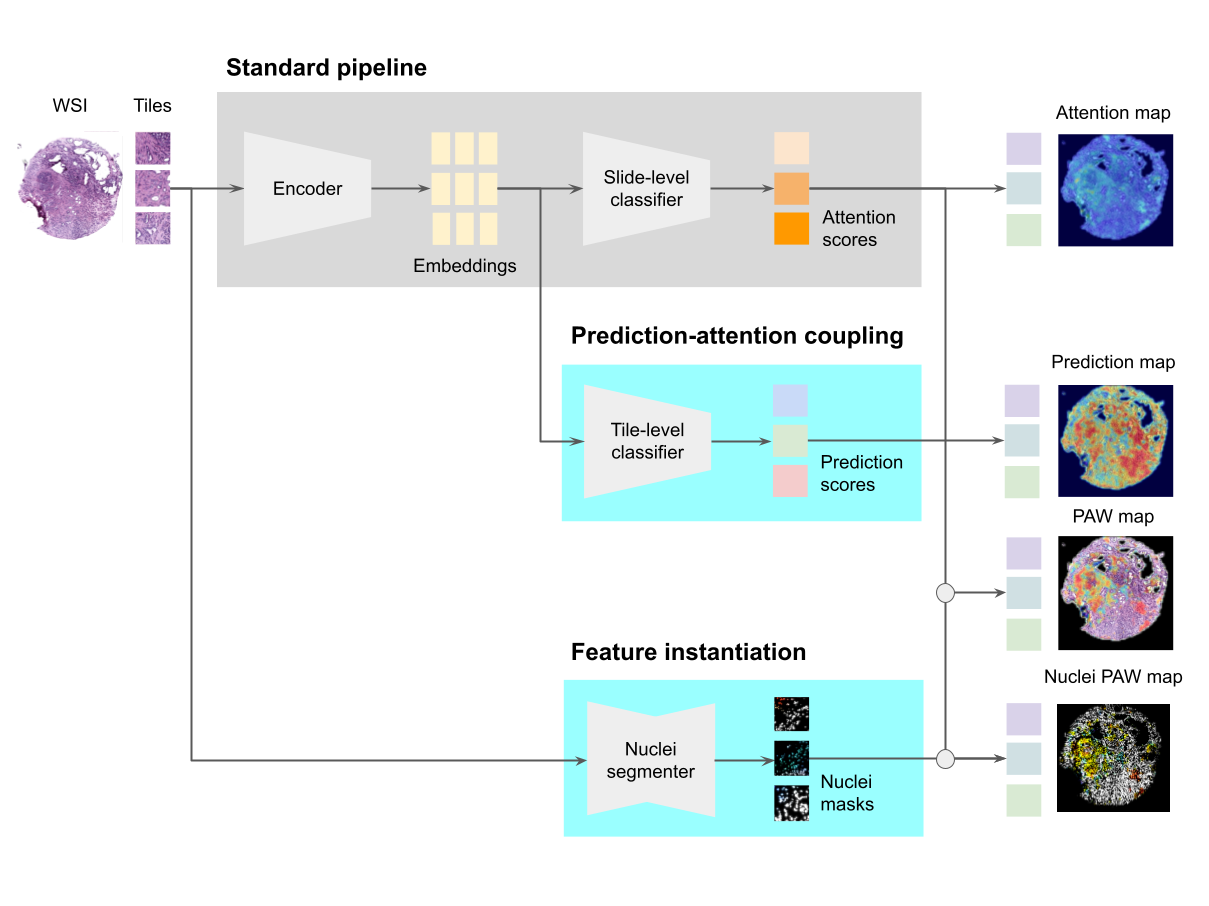}
\caption{
Overview of prediction-attention coupling and feature instantiation in multi-instance learning (MIL). 
PAW maps correspond to prediction-attention-weighted maps obtained by combining the tile-level attention and prediction scores.
The blue boxes indicate the new modules compared to the standard MIL pipeline (grey box).} \label{fig1}
\end{figure}

In this section we introduce the MIL model that we use, then we explain how to derive prediction-attention-weighted (PAW) maps to enhance the interpretability of the regions flagged by the model, and finally provide a feature instantiation pipeline to compute human interpretable features in these regions (Fig. \ref{fig1}).

\subsubsection{Self-interactive MIL:} 
Similarly with other MIL frameworks e.g. \cite{Lu2021}, we use an encoder, $f_{enc}(.)$ to generate $N_{embeds}$ tile-level embeddings, $E_{ij} \in \mathcal{R}^{N_{embeds}}$, for each of the $N_{tiles_j}$ tiles $X_{ij} \in [0,255]^{3 \times d^2}$ of sample $j$, following 
\begin{equation}
E_j = \{E_{ij} = f_{enc}(X_{ij})\}_{i=1}^{N_{tiles_j}}.
\end{equation}
Embeddings are then put through to a slide-level classifier, $f_{slide}(.)$, which predicts the label of the whole slide and generates tile-level attention scores $a_j = \{ a_{ij} \}_{i=1}^{N_{tiles_j}}$, where $a_{ij} \in [0,1]$, following 
\begin{equation}
a_j = f_{slide}(E_{j}).
\end{equation}
This model is further refined following the self-interactive (SI) learning algorithm \cite{Hu2022}, which uses the $K$ highest attention tiles of each sample $j$ to re-optimise the encoder $f_{enc}(.) \rightarrow f_{enc}^{*}(.)$ paired with a tile-level classifier $f_{tile}(.) \rightarrow f_{tile}^{*}(.)$.
The slide-level classifier is then re-optimised $f_{slide}(.) \rightarrow f_{slide}^{*}(.)$ based on the refined encoder $f_{enc}^*(.)$.
This iterative refining process produces refined tile-level embeddings $E_j \rightarrow E_{j}^{*}$ and attention scores $a_j \rightarrow a_j^{*}$, tailored to the specific task, as well as tile-level prediction scores $p^*_{ij} \in [0, 1]$, following
\begin{equation}
p_{j}^{*} = \{p_{ij}^* = f_{tile}^*(X_{ij})\}_{i=1}^{N_{tiles_j}}. 
\end{equation}

\subsubsection{PAW maps:}
We introduce prediction-attention-weighted (PAW) maps, which is a novel technique to embed the influence of each tile on the prediction, either negative or positive, into attention maps, by subsetting prediction scores of tiles based on an attention threshold $q$, 

\begin{equation}
PAW_j = \{PAW_{ij} = p_{ij}^* \}_{i \in a_j^* > q}.
\end{equation}

This allows us to subdivide regions of high attention into negative and positive evidence of the class that we are predicting. For instance, when predicting tumour/benign tissue, this allows us to separate normal from tumour tissue in high attention regions.

\subsubsection{PAW feature instantiation:}
We also introduce a feature instantiation pipeline to translate the regions of negative and positive evidence into human interpretable features. 
We illustrate this by quantifying the attention and prediction scores of individual cells, obtained by inlaying the PAW maps into nuclei segmentation masks, 
\begin{equation}
PAW'_{j} = \left\{ PAW_{ij} \times f_{nuclei}(X_{ij}) \right\}_{i=1}^{N_{tiles_j}}, 
\end{equation}
where $f_{nuclei}$ is a nuclei segmentation model \cite{Kumar2020}.
This allows us to calculate interpretable features, such as nuclei density, which corresponds to the area covered by nuclei, relative to the total tile area $d^2$, $d_{nuclei,ij} = \sum_{x}^{d} \sum_{y}^{d}f_{nuclei}(X_{ij})/d^2$, in highly-predictive regions.

\section{Experiments and results}

\begin{figure}[t]
\centering
\includegraphics[width=\textwidth]{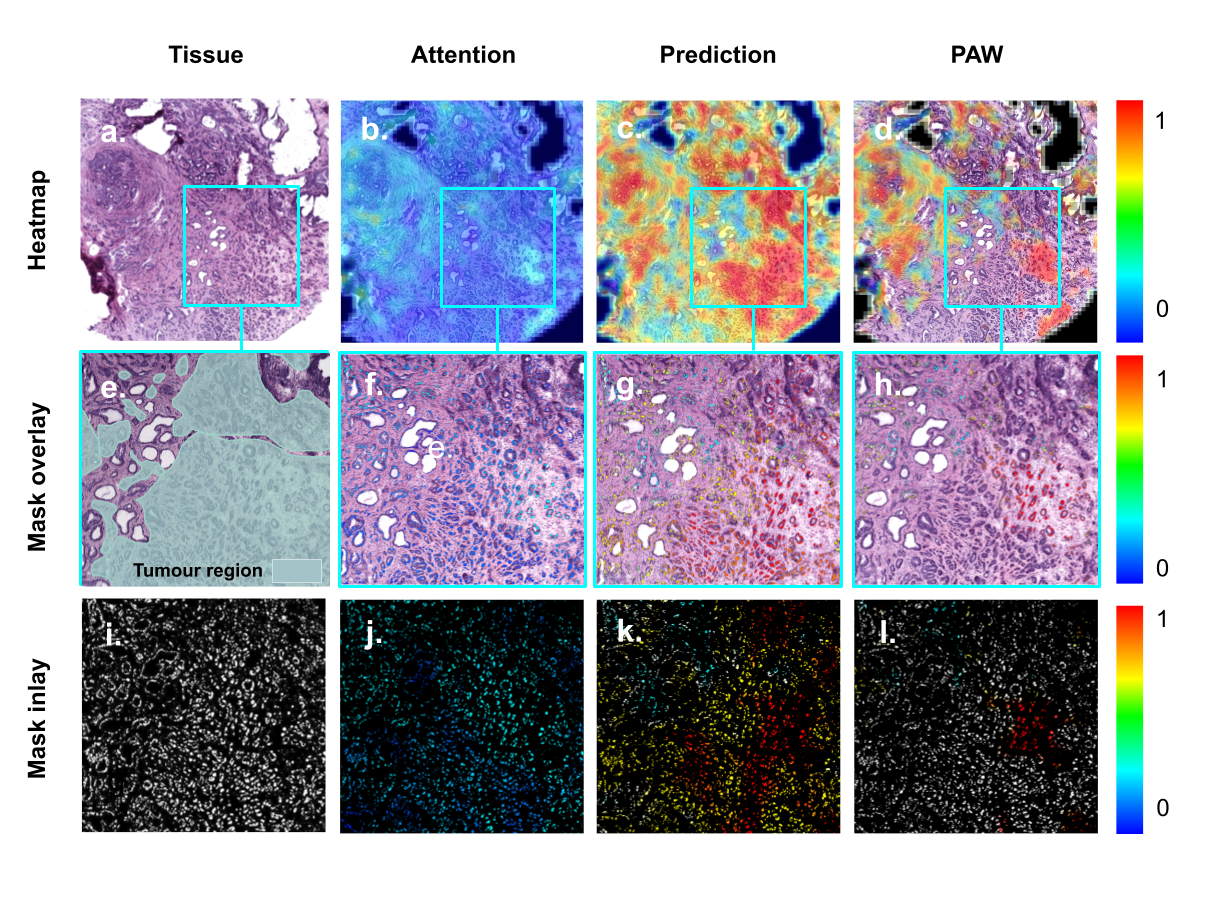}
\caption{
Attention, prediction, and prediction-attention-weighted (PAW) maps obtained from the slide- and tile-level tumour classifiers (\textbf{b}-\textbf{d}).
Colour levels, from blue to red, indicate a higher attention (\textbf{b}, \textbf{f}, and \textbf{j}) and prediction score (\textbf{c, d, g, h, k}, and \textbf{l}), comprised between 0 and 1.
Heatmaps are overlaid and inlaid with nuclei segmentation masks to better distinguish cell population structures (\textbf{e}-\textbf{h} and \textbf{i}-\textbf{l}).
The region highlighted by the light blue square corresponds to a region with more normal-looking tissue in the top-left half, and tumour tissue in the bottom-right half.
PAW maps are obtained by dividing attention-high regions into prediction-low and -high sub-regions (\textbf{d}, \textbf{h}, and \textbf{l}).
Image \textbf{e.} shows the tumour region annotated by a trained pathologist.
} 
\label{fig2}
\end{figure}

\begin{figure}[t]
\centering
\includegraphics[width=\textwidth]{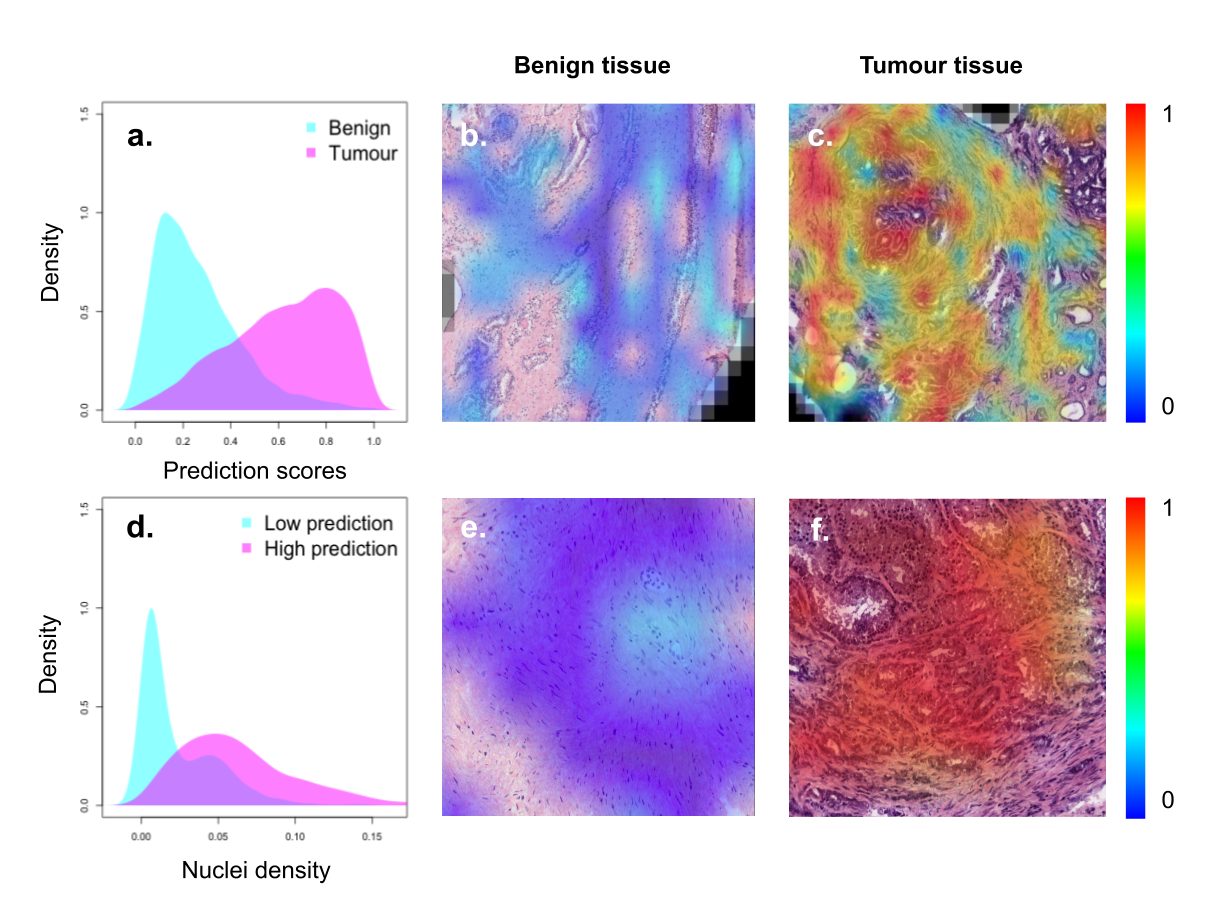}
\caption{
Comparison of tumour prediction scores (\textbf{a}), PAW maps (\textbf{b, c, e}, and \textbf{f}), and nuclei density (\textbf{d}) found in a benign and a tumour sample.
Graph \textbf{a.} shows the difference in prediction scores obtained for a benign (cyan) and tumour sample (magenta).
Graph \textbf{d.} shows the difference in nuclei density found by dividing regions of high-attention into regions of tumour-low (cyan) and -high predictions (magenta).
Images \textbf{b.} and \textbf{c.} show a high-level view of a PAW map of a benign and tumour tissue region, respectively.
Images \textbf{e.} and \textbf{f.} are a close-up of some exemplar benign and tumour regions.
} 
\label{fig3}
\end{figure}

\begin{figure}[t]
\centering
\includegraphics[width=\textwidth]{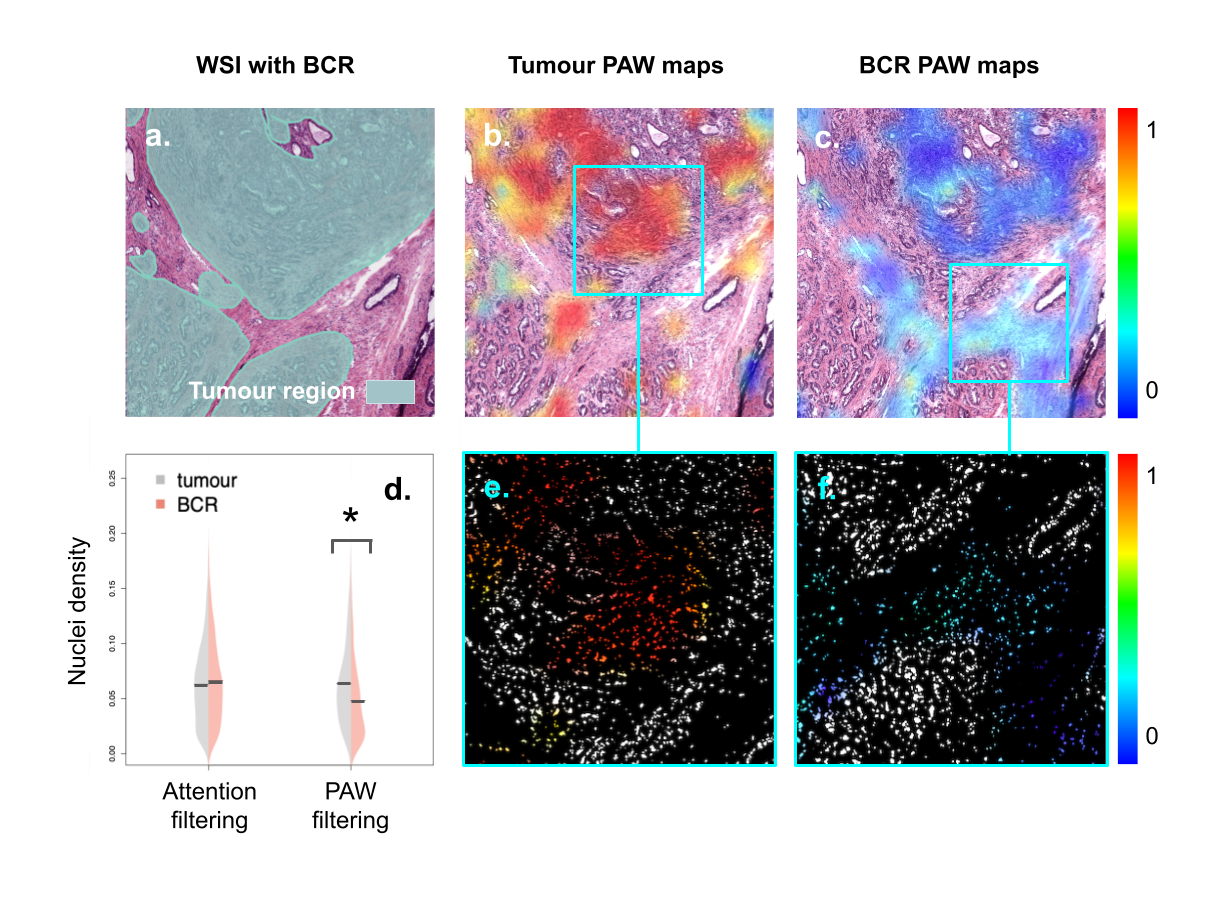}
\caption{
Comparison of exemplar PAW maps (\textbf{b} and \textbf{c}), nuclei PAW maps (\textbf{e} and \textbf{f}), and nuclei density (\textbf{d}) obtained by applying the tumour and BCR classifiers to a BCR sample.
Image \textbf{a.} shows the tumour region annotated by a trained pathologist.
Graph \textbf{b.} and \textbf{c.} display tumour and BCR prediction scores, respectively, in regions of high attention.
Colours, from blue to red, correspond to low to high prediction scores.
Graph \textbf{d.} compares the nuclei density in high-attention regions with high tumour and BCR prediction scores (PAW filtering) vs only looking at high-attention regions (attention filtering).
Significance level of $p < 0.01$ obtained from pairwise student t-tests are shown by an asterisk.
Graph \textbf{e.} and \textbf{f.} show the spatial structures of cell populations in tumour-high and BCR-high regions, respectively.
} 
\label{fig4}
\end{figure}

In order to demonstrate the usefulness of prediction-attention coupling in MIL, we characterised tissue regions and features (i.e. nuclei density) that are predictive of diagnosis and prognosis.
Diagnosis corresponds to the detection of malignant tumour tissue in a sample \cite{Humphrey2004}, while prognosis corresponds to the outcome of the treatment, established by biochemical recurrence (BCR) \cite{Pinckaers2022}.
Tumour regions are expected to feature a higher nuclei density, as a result of tumour cell proliferation, than benign regions \cite{Gibbs2009}.
Given that BCR is thought to associate with more aggressive tumour forms \cite{Pinckaers2022}, we expect BCR-high regions to be located within tumour-high regions and hence also feature high nuclei density.

\subsubsection{Dataset:}
The dataset we used for training and testing is taken from the international cancer genome consortium (ICGC UK Prostate Group).
This corresponds to over 600 H\&E stained whole slide images, obtained from 95 patients.
381/516 samples were tissue diagnosed with malignant tumour, and 98/663 samples came from patients with BCR.
Whole slide images were divided into 4 overlapping sets of 256$\times$256 pixels wide tiles.

\subsubsection{Training and performance:}
We independently trained two Inter-MIL models \cite{Hu2022}, one to predict tumour, and the other to predict BCR.
We use the SI learning algorithm to train the encoder, tile- and slide-level classifier \cite{Hu2022}, with $N_{embeds} = 32$ features for the embeddings, $K = 16$ highest-attention tiles to refine the encoder, $N_{ensemble} = 30$ ensemble elements for the slide-level classifier, three rounds of refining with $N_{epochs} = 20$ refining epochs for the encoder and tile-level classifier and $N_{epochs} = 80$ for refining the slide-level classifier, a two-third/one-third training/testing split, a convolutional neural network (CNN) architecture for the encoder, and a suite of fully connected layers for the slide- and tile-level classifiers (see \cite{Hu2022} for details). 
Training took 5 hours for each model in Python v3.10.8 and Pytorch v1.13.0 on MacBook pro M1 MAX 2022, and results could be reproduced by repeat training.
The performance of the slide-level classifier reached $AUC_{train} = 0.96 \pm 0.01$ and $AUC_{test} = 0.94 \pm 0.04$ for tumour diagnosis, and $AUC_{train} = 0.78 \pm 0.04$ and $AUC_{test} = 0.68 \pm 0.05$ for BCR prediction.

\subsubsection{PAW maps:}
We computed and visualised PAW maps, focusing on the tumour classification task for the sake of conciseness (Fig. \ref{fig2}).
The PAW maps show tumour prediction scores in high attention regions, that is $a_{j} \geq q_{0.75}(a_j)$, where $q_{0.75}$ denotes the upper quartile of the attention distribution of sample $j$.
We find that both the attention and prediction maps identify the tumour areas (Fig. \ref{fig2}, b and c), and that PAW maps allow us to further refine high attention regions into sub-regions that are tumour-low and tumour-high (Fig. \ref{fig2}d).
The visualisation of individual nuclei clearly shows the divide between more benign cells in the top-left half of the image and tumour cells in the bottom-right half (Fig. \ref{fig2}, g and k).
The PAW map combined with the nuclei mask points at the specific cell population that contributes most (and positively in this case) to diagnosis in this region (Fig. \ref{fig2}, h and l).

\subsubsection{PAW feature instantiation:}
We demonstrate here how these maps can be used for feature instantiation.
For the sake of simplicity, we focus on nuclei density, measured by the area covered by nuclei relative to the total area of the tile.
We compare the tumour prediction scores obtained in a benign and a tumour sample (Fig. \ref{fig3}a), as well as the nuclei density of high-attention regions with a low and high tumour prediction score (Fig. \ref{fig3}d).
Expectedly, tiles of the benign sample receive lower prediction scores than the tumour sample (Fig. \ref{fig3}a, b, c, e, and f).
Also in line with our expectations, regions with high tumour prediction scores tend to have a higher nuclei density than more benign tiles (Fig. \ref{fig3}d) \cite{Gibbs2009}.

\subsubsection{Comparing diagnosis and prognosis:} 
Finally, we compared PAW maps and features obtained from the tumour and BCR classifiers in order to determine whether regions important for diagnosis and prognosis overlapped and presented similar features.
For the sake of conciseness, we only present the within-sample analysis of a single tissue sample from a verified BCR case.
In this case, we find that tumour-associated regions do not obviously overlap with BCR-associated regions, as most tumour-high prediction scores  are found in the tumour area in the top of the image (Fig. \ref{fig4}b), whereas BCR-high prediction scores are found in the stromal region adjacent to the tumour region at the bottom of the image (Fig. \ref{fig4}c).
Furthermore, we find that nuclei density tends to be lower in BCR-predictive regions, compared to tumour-predictive regions (Fig. \ref{fig4}d, PAW filtering), and that this is not revealed by looking at attention-high regions only (Fig. \ref{fig4}d, attention filtering).
Overall, this suggests that BCR features may not be as clearly tumour-associated as previously thought \cite{Finak2008,Wu2016,Kemi2018,Gonzalez2022}.

\section{Discussion}

In this work, we introduced the novel concept of prediction-attention coupling to the multiple instance learning (MIL) framework. 
We demonstrate how this concept can be used to derive prediction-attention-weighted (PAW) maps, allowing us to further divide regions of high attention into sub-regions that are negatively or positively associated with the class labels.
Secondly, we illustrate how this could guide biological feature instantiation, by computing nuclei density in regions identified as highly-predictive by the PAW maps.
Our approach holds the promise of providing additional information for diagnosis and prognosis that is otherwise not available. Our data indicates that tumour-predictive regions identified with PAW maps tended to feature a higher nuclei density, and that BCR-predictive regions tended to locate in regions adjacent to tumour, in areas featuring lower cell density.
This suggests that regions predictive of prognosis and diagnosis may not overlap entirely, indicating that further investigations into tumour-adjacent regions could improve our capacity to predict response to treatment.
Overall, our work is a crucial step towards making MIL models more interpretable, and though we only presented results on nuclei segmentation, our approach can include multiple additional channels of information, such as additional segmentation masks (e.g. prostatic glands), spatial transcriptomics, or histo-immuno chemistry, to dive deeper into biological mechanisms.


\subsubsection{Acknowledgments.}
Additional to the named authors, the CRUK ICGC Prostate Group also contains the following members: 
Adam Lambert, University of Oxford, Oxford, UK; Anne Babbage, Hutchison/MRC Research Centre, Cambridge University, Cambridge, UK
Claudia Buhigas, Norwich Medical School, University of East Anglia, Norwich, UK;
Dan Berney, Department of Molecular Oncology, Barts Cancer Centre, London, UK;
Nening Dennis, Royal Marsden NHS Foundation Trust, London and Sutton, UK;
Sue Merson, The Institute Of Cancer Research, London, UK;
Alastair D. Lamb, Nuffield Department of Surgical Sciences, University of Oxford, Oxford, UK;
Adam Butler, Cancer Genome Project, Wellcome Trust Sanger Institute, Hinxton, UK;
Anne Y. Warren, Department of Histopathology, Cambridge University Hospitals NHS Foundation Trust, Cambridge, UK;
Vincent Gnanapragasam, Department of Surgical Oncology, University of Cambridge, Addenbrooke’s Hospital, Cambridge, UK;
G. Steven Bova, Prostate Cancer Research Center, Faculty of Medicine and Health Technology, Tampere University, Finland;
Christopher S. Foster, HCA Laboratories, London, UK;
David E. Neal, Department of Surgical Oncology, University of Cambridge, Addenbrooke’s Hospital, Cambridge, UK;
Yong-Jie Lu, Centre for Molecular Oncology, Barts Cancer Institute, Queen Mary University of London, London, UK;
Zsofia Kote-Jarai, The Institute of Cancer Research, London, UK;
Robert G. Bristow, Manchester Cancer Research Centre, Manchester, UK;
Andy G. Lynch, School of Medicine/School of Mathematics and Statistics, University of St Andrews, St Andrews, UK;
Daniel S. Brewer, Norwich Medical School, University of East Anglia, Norwich, UK;
David C. Wedge, Manchester Cancer Research Centre, Manchester, UK;
Rosalind A. Eeles, The Institute of Cancer Research, London, UK;
Colin S. Cooper, Norwich Medical School, University of East Anglia, Norwich, UK. 
Finally, we thank the Rittscher and Mills lab for insightful feedback on early versions of the work. 
This work was supported by grant MA-ETNA19-005 'Major Awards, Existing Trials: New Answers' funded by Prostate Cancer UK.

\subsubsection{Statement of authorship.}
Willem Bonnaff\'e designed the methodology, performed the analysis, wrote the manuscript. 
The CRUK ICGC Prostate Group provided the dataset.
Freddie Hamdy co-directed the work.
Yang Hu helped designing the methodology, contributed to the manuscript.
Ian Mills co-directed the work, provided feedback on the manuscript.
Jens Rittscher co-directed the work, contributed to the manuscript.
Clare Verrill provided annotations and feedback on results.
Dan J. Woodcock co-directed the work, provided feedback on the manuscript.

\subsubsection{Data accessibility.}
The dataset and code used for the analysis can be accessed upon reasonable request.

%
%
%
\bibliographystyle{splncs04}
\bibliography{export}

\end{document}